\documentclass{PoS}
\usepackage[FIGBOTCAP,TABBOTCAP,tight]{subfigure}
\usepackage{tikz}
\usepackage{graphicx}
\usepackage{grffile}
\usepackage{overpic}
\usepackage[utf8]{inputenc}
\usepackage{amsmath}
\usepackage{amssymb}

\graphicspath{{./fig/}}

\title{Gauge Corrections to Strong Coupling Lattice QCD on Anisotropic Lattices
}

\ShortTitle{Gauge Correctinos on Anisotropic Lattices}

\author{Jangho Kim\\
        Institut f{\"u}r Theoretische Physik, Goethe-Universit{\"a}t Frankfurt am Main, Max-von-Laue-Str. 1, 60438 Frankfurt am Main, Germany\\
        E-mail: \email{jkim@th.physik.uni-frankfurt.de}}

\author{Marc Klegrewe\\
        Fakult{\"a}t f{\"u}r Physik, Universit{\"a}t Bielefeld, Universit{\"a}tstasse 25, D33619 Bielefeld, Germany\\
        E-mail: \email{mklegrewe@physik.uni-bielefeld.de}}    

\author{\speaker{Wolfgang Unger}\\
        Fakult{\"a}t f{\"u}r Physik, Universit{\"a}t Bielefeld, Universit{\"a}tstasse 25, D33619 Bielefeld, Germany\\
        E-mail: \email{wunger@physik.uni-bielefeld.de}}

\abstract{
Lattice QCD with staggered fermions can be formulated in dual variables to address the finite baryon
density sign problem. In the past we have performed simulations in the strong coupling regime,
including leading order gauge corrections. In order to vary the temperature for fixed $\beta$ it was
necessary to introduce a bare anisotropy. In this talk we will extend our work to include results from
a non-perturbative determination of the physical anisotropy $a_\sigma$/$a_\tau = \xi(\gamma, \beta)$, which is necessary to
unambiguously locate the critical end point and the first order line of the chiral transition.
}

\FullConference{37th International Symposium on Lattice Field Theory - Lattice2019\\
		16-22 June 2019\\
		Wuhan, China}

\def \as {{a_{\sigma}}}
\def \at {{a_{\tau}}}
\def \Ns {{N_{\sigma}}}
\def \Nt {{N_{\tau}}}
\def \Nc {{N_{c}}}

\begin{document}

\section{Introduction}

Despite many attempts and partial successes to address the finite density sign problem in lattice QCD, a solution applicable to the full parameter space (temperature $T$, baryon chemical potential $\mu_B$, quark mass $m_q$ and lattice gauge coupling $\beta$) has not yet been established.
Here we report on the incremental progress to unravel the phase diagram in the strong coupling regime of lattice QCD with staggered fermions, based on a leading order strong coupling expansion valid to $\mathcal{O}(\beta)$ \cite{deForcrand2014,Gagliardi2017,Kim2019}. The recent progress to address higher order corrections \cite{Gagliardi2019} are not yet considered in full Monte Carlo simulations.

The phase diagram of lattice QCD in the strong coupling limit has been investigated since more than 30 years \cite{Kawamoto1981,Rossi1984,Karsch1989,Kawamoto2005} and is by now well known, with the Worm algorithm as a main Monte Carlo tool to investigate its features \cite{Adams2003,Forcrand2010,Unger2011}. Beyond the strong coupling limit, the leading order gauge corrections have been included as well, but ambiguities on the phase boundary arising when using different $\Nt$ have not yet been addressed. These ambiguities have so far only been successfully resolved in the strong coupling limit (both in the chiral limit \cite{deForcrand2017} and at finite quark mass \cite{Bollweg2018}).

The long-term goal is to extend the validity of the strong coupling expansion to answer an important question on the existence of the critical end point (CEP):
At strong coupling, the CEP has been located at $(a\mu_B^c,aT^c)=(1.56(3),0.80(2))$ in the chiral limit (where the CEP turns into a tri-critical point TCP), and its quark mass dependence has been investigated, with tri-critical scaling $\propto m_q^{2/5}$ for small quark masses \cite{Kim2016}. The dependence of the location of the CEP as a function of $\beta$ has not yet been determined. Whether the CEP also exists in the continuum limit remains an open question. First hints can be obtained by monitoring the $\beta$-dependence of the CEP for small $\beta$: if it moves to smaller $\mu_B$ (and if this behaviour is monotonous), it may exist; if it moves to larger $\mu_B$, it may even vanish in the continuum limit and the chiral transition is for all values of $\mu_B$ just a crossover.

The main difficulty when mapping out the phase diagram is that we need to introduce a bare anisotropy $\gamma$ in the strong coupling regime in order to vary the temperature continuously at fixed values of $\beta$. The temperature and chemical potential are however determined by the physical anisotropy $\xi\equiv \frac{\as}{\at}$, which depends non-perturbatively on $\gamma$ and the lattice gauge coupling $\beta$. 
Here we will report on how the $\beta$-dependence of $\xi$ is determined, and present preliminary results when applied to the phase diagram in the strong coupling regime.\\

\section{Dual formulation of lattice QCD}

The strong coupling regime of lattice QCD can be formulated in a dual representation and it was generalized recently to include in principle any order in $\beta$ \cite{Gagliardi2019}. In this proceedings however, we only incorporate the leading order gauge correction $\mathcal{O}(\beta)$ as outlined in \cite{Gagliardi2017} and re-derived in the appendix of \cite{Gagliardi2019}.
It is based on a series expansion in terms of the (anti-) quark hopping $\bar{d}_\mu(x)$ from the staggered Dirac operator, and plaquette occupation numbers $n_p, \bar{n}_p$ on plaquette coordinates $p=(x,\mu,\nu)$ from the Wilson gauge action. 
In contrast to previous formulations of the dual partition sum, we now adopt the notation:
\begin{align}
k_\mu(x) &= \min\left\{ d_\mu(x),\bar{d}_\mu(x) \right\},& f_\mu(x) &= d_\mu(x) - \bar{d}_\mu(x),
\end{align}
where $k_\mu(x)\in \{0,\ldots \Nc\}$ is the \textit{dimer number} and $f_\mu(x)\in \{-\Nc,\ldots \Nc\}$ is the \textit{net quark flux}. The $k_\mu(x)$ are always quark-antiquark combinations, and color singlets formed by a quark and gluon are no longer regarded as dimers (in contrast to our previous formulation - the new convention is advantageous when higher order corrections are considered).
The dual degrees of freedom $\{k, f, m, \bar{n}, n \}$ fulfill the gauge constraint at each link:
\begin{align}
f_{\mu}(x)&+{\displaystyle \sum_{\nu>\mu}}\bigg[\delta n_{\mu,\nu}(x)-\delta n_{\mu,\nu}(x-\nu)\bigg] - {\displaystyle \sum_{\nu < \mu}}\bigg[ \mu\leftrightarrow\nu\bigg] = \Nc\, q_\mu(x),
\label{GaugeConstraint}
\end{align} \\
where for the $\mathcal{O}(\beta)$ partition function, $q_{\mu}(x)\in\{-1,0, 1\}$ and $\delta n_{\mu,\nu}(x)\equiv\delta n_{p} = n_{p} -\bar{n}_{p}\in \{-1,0,1\}$.  The Grassmann constraint at each lattice site is:
\begin{align}
m_{x}+ \sum_{\pm \mu}\left( k_{\mu}(x) +\frac{|f_{\mu}(x)|}{2}\right) &= \Nc, & {\displaystyle \sum_{\pm \mu}} f_{\mu}(x) &= 0.
\label{GrassmanConstraint}
\end{align}

In terms of the above dual variables, and including a bare anisotropy $\gamma$, the partition function can be rewritten as:
\begin{align}
Z(\beta,\gamma,\mu_{q},\hat{m}_q) &=\hspace{-3mm} 
 \sum_{C=\{n_{p},\bar{n}_{p},k_{\ell},f_{\ell},m_{x} \}}\hspace{-3mm}
\sigma(C)
 \prod_{p}\frac{\tilde{\beta}^{n_{p}+\bar{n}_{p}}}{n_{p}!\bar{n}_{p}!}
 \prod_{\ell=(x,\mu)}\frac{e^{\mu_{q}\delta_{\mu,0}f_\mu(x)}\gamma^{\delta_{\mu,0}\left(|f_\mu(x)|+2k_\mu(x)\right)}}{k_{\ell}!(k_{\ell}+|f_{\ell}|)!} 
 \prod_{x}\frac{(2\hat{m}_q)^{m_{x}}}{m_{x}!}T_{i}(C_x)
 \label{DualizedPartitionFunction}
 \end{align}
with $\tilde{\beta}=\frac{\beta}{2\Nc}$, the quark chemical potential $\mu_q=\frac{1}{\Nc}\mu_B$. The three non-trivial vertex weights
\begin{align}
T_{1}&=\frac{\Nc!}{\sqrt{\Nc}},&
T_{2}&=(\Nc-1)!,&
T_{3}&=\frac{\Nc!}{\sqrt{\Nc}}&
\end{align} 
depend on the local degrees of freedom $C_x=\{m_x,k_\mu(x),f_\mu(x),n_{\mu\nu}(x),\bar{n}_{\mu\nu}(x)\}$ and are employed whenever some $n_{\mu\nu}(x)>0$ 
($\bar{n}_{\mu\nu}(x)>0$)
 and some $f_\mu(x)>1$. For $\Nc=3$, the sign
\begin{align}
\sigma(C)&=\prod_{\ell_1}\sigma(\ell_1)\prod_{\ell_3}\sigma(\ell_3),& \sigma(\ell)&=(-1)^{1+w(\ell)+N_{-}{(\ell)}}\prod_{\tilde{\ell}}\eta_\mu(x)
\end{align}
factorizes into single fermion ($|f_\mu(x)|=1$) and triple fermion loops ($|f_\mu(x)|=3$). This factorization no longer holds beyond $\mathcal{O}(\beta)$, see \cite{Gagliardi2019}. 
The dual degrees of freedom are color singlets which are no longer just baryons and mesons as in the strong coupling limit: the gauge corrections will resolve the quark structure of the point-like baryons and mesons, making them effectively spread out over one or more lattice spacings. 
The reason why the sign problem is mild in the strong coupling limit is that baryons are heavy, where $\Delta_f\simeq 10^{-5}$. 
This is still approximately true for $\beta \lesssim 1$, where the sign problem remains manageable. For details see \cite{Kim2019}.

In the following we will consider the chiral limit of the partition function Eq.~(\ref{DualizedPartitionFunction}), which implies $m_x=0$ and which has the symmetry group : 
 \begin{align}
 &U(1)_V\times U(1)_{55}:&\chi(x)&\mapsto e^{i\epsilon(x)\theta_A+i\theta_V}\chi(x),& \epsilon(x)&=(-1)^{x_1+x_2+x_3+x_4},
 \end{align}
 with $U(1)_V$ the baryon number conservation and $U(1)_{55}$ the remnant chiral symmetry which is broken spontaneously at low temperatures and densities. 
In Sec.~\ref{ChiralTransition} we will address the chiral critical line that terminates in a tri-critical point before turning first order.

\newcommand{\mh}{\hat{m}}

\section{Anisotropy Calibration at finite $\beta$}

It is crucial to understand the relationship between the bare anisotropy $\gamma$ and the non-perturbative anisotropy $\xi\equiv \frac{\as}{\at}$ (with $a\equiv \as$ the spatial and $\at$ the temporal lattice spacing) in order set the temperature and chemical potential consistently for various $\Nt$. Anisotropic lattices are necessary in the strong coupling regime since at fixed $\beta$ this is the only way to vary the temperature continuously \cite{Levkova2006,Klegrewe2018}. 
The precise correspondence between $\xi$ and $\gamma$ has been established in the strong coupling limit and in the chiral limit \cite{deForcrand2017}, resulting in 
\begin{align}
\xi(\gamma)&\approx \kappa\gamma^2+\frac{\gamma^2}{1+\lambda\gamma^4},& \kappa&=0.781(1),
\end{align}
and at finite quark mass in \cite{Bollweg2018}, where it was shown that 
$\kappa(m_q)=\lim\limits_{\xi\rightarrow \infty}\frac{\xi}{\gamma^2}$ has a simple mass dependence in the strong coupling limit.
The basic idea of the anisotropy calibration is to identify a conserved current and scan in $\gamma$ such that the lattice is physically isotropic for a fixed aspect ratio:
\begin{align}
\Ns \as&\stackrel{!}{=} \Nt \at & \Rightarrow && \xi&=\frac{\Nt}{\Ns}.
\end{align}
 The conserved current is related to the pion \cite{Chandrasekharan2003}
\begin{align}
 j_\mu(x)&=\epsilon(x)\left(k_\mu(x)-\frac{1}{2}|f_{\mu}(x)|\right),
 \label{current}
 \end{align}
 with $\epsilon(x)=\pm 1$ the parity of site $x$. Eq.~(\ref{current}) is the generalization of the strong coupling limit (where $f_{\mu}(x)\in \{-\Nc,0,\Nc\}$ is the baryon flux through that link) to incorporate gauge corrections.
This allows us to extend the anisotropy calibration to finite $\beta$ to obtain $\xi(\gamma,\beta)$.
Away from the strong coupling limit it is in principle necessary to include a second bare anisotropy $\gamma_G$ in the gauge part
\begin{align}
 \beta^{n_{p}+\bar{n}_{p}}\quad &\rightarrow \quad\beta_{\sigma}^{n_{p_{\sigma}}+\bar{n}_{p_{\sigma}}}\beta_{\tau}^{n_{p_{\tau}}+\bar{n}_{p_{\tau}}},& \gamma_G&=\sqrt{\frac{\beta_\tau}{\beta_\sigma}}
\end{align}
and then scan in both the fermionic and gauge anisotropy to obtain $\xi(\gamma,\gamma_G,\beta)$. On finer lattices this is indeed necessary \cite{Karsch1989}, but in the strong coupling regime, where we cannot set a scale, it is an unnecessary complication: as $\beta$ is increased, the lattices needed to study the chiral phase transition will eventually become isotropic, and beyond this point, the temperature is varied via $a(\beta)$. 
In this proceedings, we will always set $\gamma_G=1$ and leave the more general setup for the future.

\begin{figure}[h!]
\centerline{
\includegraphics[width=0.45\textwidth]{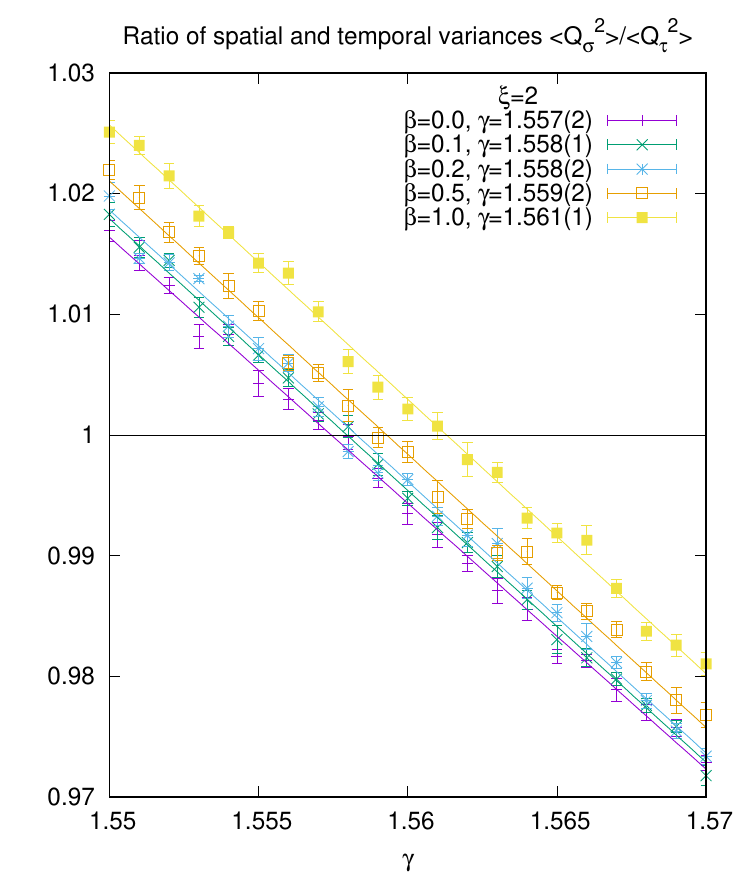}
\includegraphics[width=0.45\textwidth]{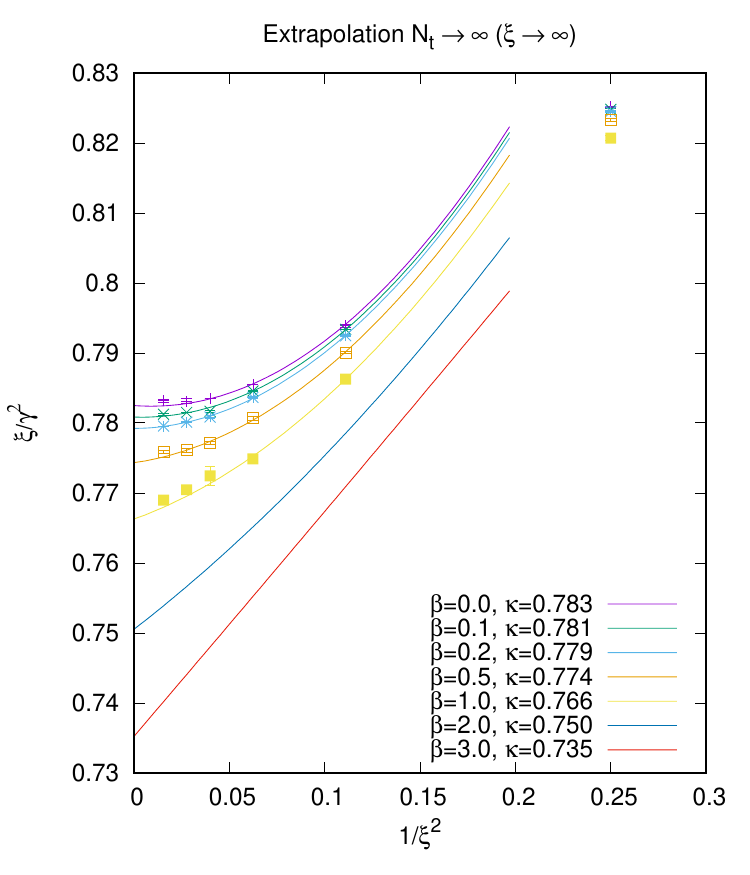}
}
\caption{
 \emph{Left:} Determination of $\gamma$ for various $\beta$ by requiring the ratio of charge fluctuations to be equal, shown for $\xi=2$. \emph{Right:} Extrapolation of the correction factor $\xi/\gamma^2$ towards continuous time to yield $\kappa(\beta)$.
}
\label{Callibration}
\end{figure}

In Fig.~\ref{Callibration} (\emph{left}) we show the anisotropy calibration for various fixed $\beta$: On lattices $\Ns^3\times \Nt$ with aspect ratios $\xi=2,3,4,5,6,8$ we obtain the value of $\gamma(\xi)$ 
where the ratio of the temporal and spatial fluctuations of the conserved charge $Q_t$, $Q_s$ are equal. This is repeated for various $\beta$. 
Since the partition function Eq.~(\ref{DualizedPartitionFunction}) depends on $\gamma$ and $\Nt$, the bare (mean field) temperature $[aT]_{\rm mf}=\frac{\gamma^2}{\Nt}$ needs to be corrected by the non-perturbative factor $[\xi/\gamma^2]_\beta$, shown in Fig.~\ref{Callibration} (\emph{right}), to yield the correct temperature $aT=\frac{\xi(\gamma)}{\Nt}$.  Our result allows to define the Euclidean continuous time limit $\at\rightarrow 0$ unambiguously at fixed $\beta$: the  temperature and chemical potential are then defined as 
\begin{align}
 aT&=\kappa(\beta)[aT]_{\rm mf},& a\mu_B&=\kappa(\beta)[a\mu_B]_{\rm mf}& \text{with}&& \kappa(\beta)&=\lim\limits_{\xi\rightarrow \infty} [\xi/\gamma^2]_\beta.
\label{Renormalization}
 \end{align}

\section{Gauge Corrections to the Phase Diagram and Density of States}
\label{ChiralTransition}

\begin{figure}
\centerline{\includegraphics[width=0.9\textwidth]{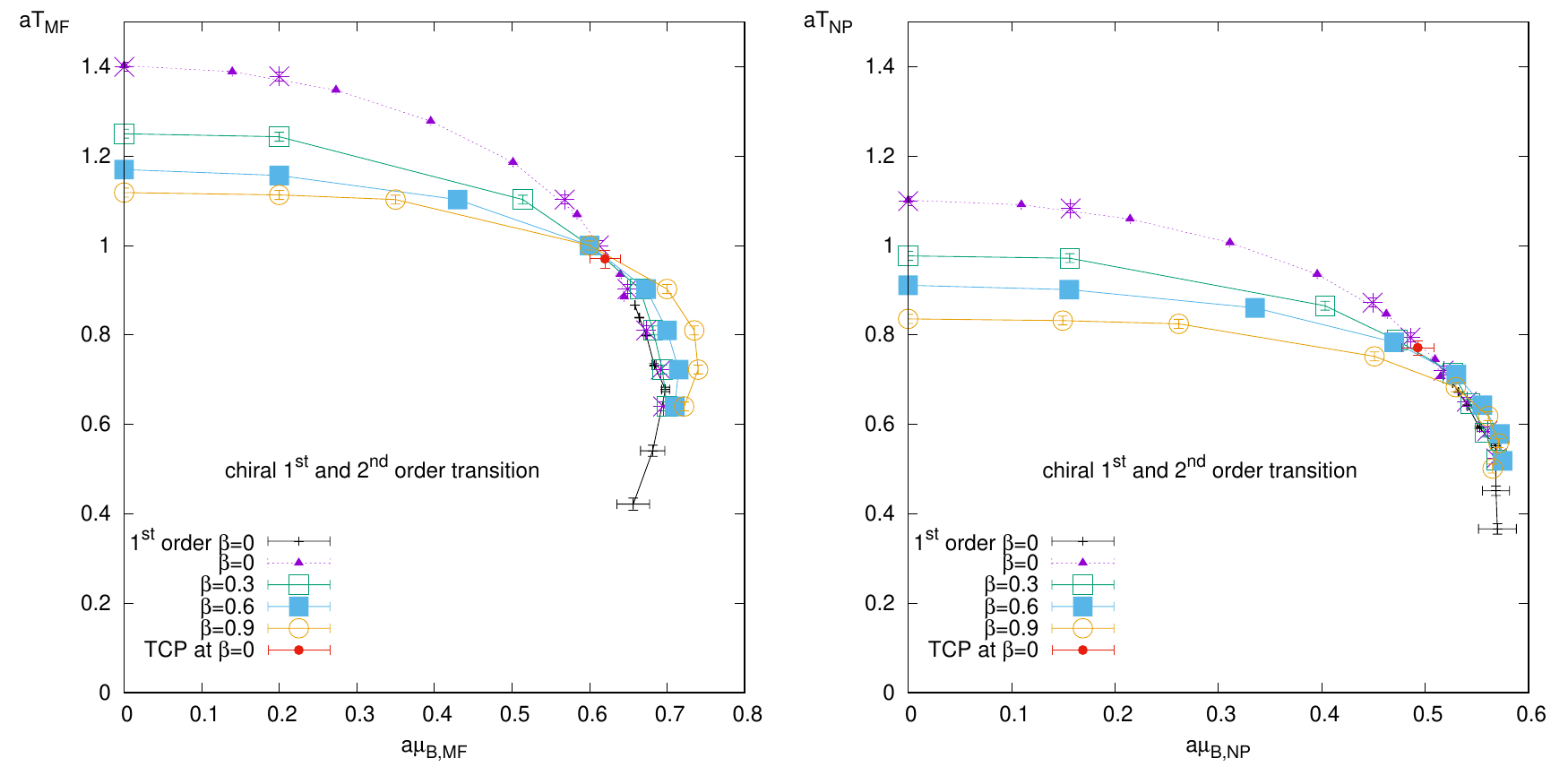}}
\caption{Comparison of the phase boundary with the mean field definition of the temperature \emph{(left)} and its non-perturbative counterpart \emph{(right)}, resulting in a collapse of the first order line for all values of $\beta$ considered.
}
\label{phasediag}
\end{figure}

We will now focus on a particularly important application of the previous result: the modification of the chiral transition within the grand-canonical phase diagram, when taking into account the non-perturbative definition of temperature and chemical potential Eq.~(\ref{Renormalization}).
In Fig.~\ref{phasediag} we show the effect of applying the $\beta$-dependent correction factor $[\xi/\gamma^2]_\beta$ to the phase boundary, for the various $\beta$ in a regime where the sign problem is manageable. All data have been measured via the Worm algorithm in combination with plaquette updates, on lattices $\Ns^3\times 4$  We observe that the back-bending at lower temperatures vanishes. 
This behaviour meets our expectations, but we require larger lattices and should check that we have the same finding also on lattices with $\Nt>4$.

\newcommand{\D}{{\mathcal{D}_b}}
\newcommand{\E}{{\mathcal{E}_{a-b}}}
\newcommand{\DD}[1]{{\mathcal{D}_{#1}}}
\newcommand{\EE}[1]{{\mathcal{E}_{#1}}}

\newcommand{\iikk}{_{\bm{i\,\raisemath{3pt}{i},k\,\raisemath{3pt}{k}}}}

\begin{figure}
\includegraphics[width=\textwidth]{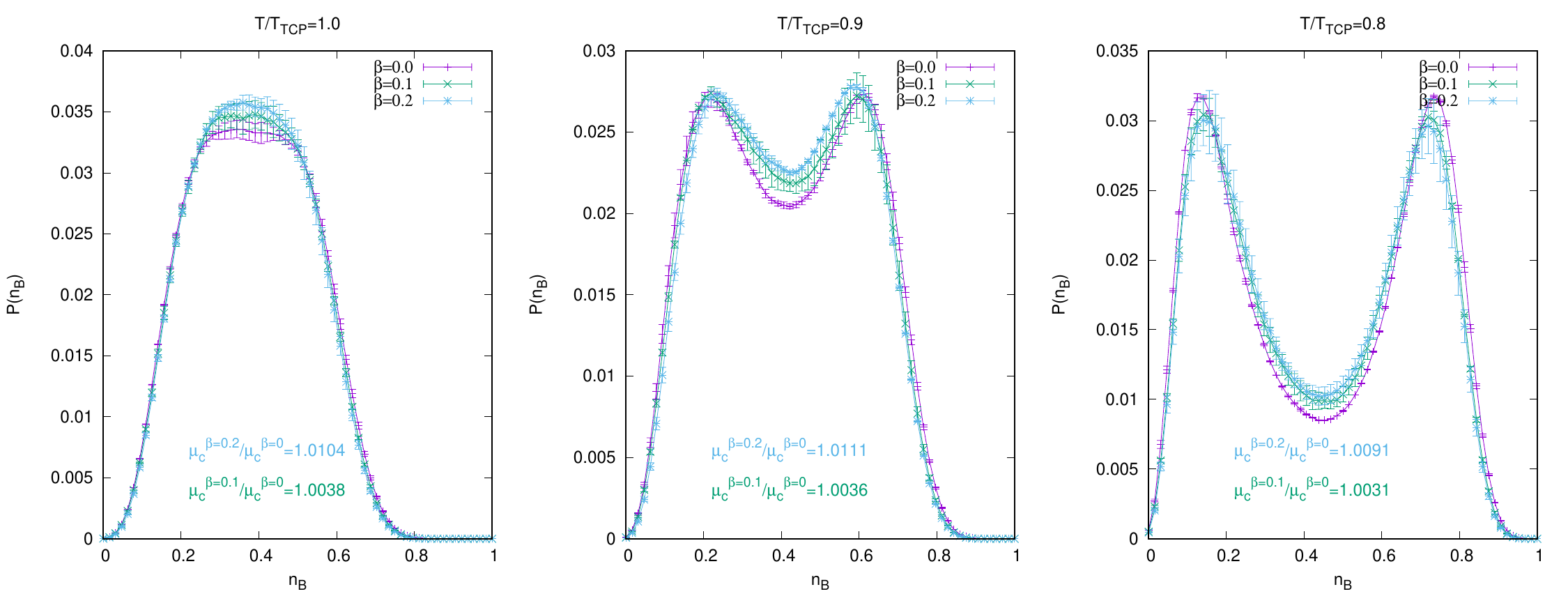}
\caption{The density of states weighted to the critical chemical potential $\mu_B^c$, showing a double peak structure for $aT<aT_{TCP}$. The value of
$\mu_B^{1^{\rm st}}$ only very mildly grows with $\beta$.}
\label{DensityStates}
\end{figure}

We also investigate the density of states on anisotropic lattices, which can be measured via the Wang-Landau method. Since the quark fluxes $f_\mu(x)$ form world lines, and the total number of quark fluxes wrapping around in temporal direction is a multiple of $\Nc$ due to the gauge constraint Eq.~(\ref{GaugeConstraint}), it is possible to define baryon number sectors $N_B\in\{-\Ns^3,\ldots, \Ns^3\}$ and allow updates that modify the baryon number by one unit. We will explain the details of the canonical simulations 
and the resulting canonical phase diagram is in the $n_B -T$ plane in a forthcoming publication. The analysis of the  density of states in $N_B$ as shown in Fig.~\ref{DensityStates} can yield additional insights concerning the first order phase boundary below the TCP: the density of states is weighted with $e^{N_B \mu_B/T }$ for various $\beta$ to the critical chemical potential $\mu_B^{1^{\rm st}}$, where the peak heights are equal. We observe that the first order transition weakens with $\beta$, and that the the critical chemical potential $\mu_B^{1^{\rm st}}$ increases only slightly with $\beta$. This is in agreement with the findings of the $\beta$-dependence of the nuclear transition at low temperatures on isotropic lattices \cite{Kim2019}.

\section{Conclusions}

We determined the non-perturbative relation between the bare anisotropy $\gamma$ and the lattice anisotropy $\xi=\frac{a}{a_t}$ at finite $\beta$ in the range of validity $\beta\leq 1$, based on the leading order partition function. The results have been used to define the temperature and baryon chemical potential unambiguously. The extrapolation $a_t\rightarrow 0$ is under control.
This may even allow to extend the existing Monte Carlo simulations in Euclidean continuous time to finite $\beta$ in the future.

 The main (still preliminary) finding on the phase boundary of lattice QCD in in the chiral limit is that the first order line is not $\beta$-dependent after the non-perturbative correction of the temperature and chemical potential. This is consistent with mean-field theory \cite{Miura2016} and results on isotropic lattices. Whether the first order line is $\beta$-dependent for $\beta>1$ and whether the tri-critical point moves to larger or smaller chemical potential when $\beta$ is increased
 requires further investigation. Most likely higher order corrections need to be included, as outlined in \cite{Gagliardi2019}.

We have also presented first results on the $\beta$-dependence of the density of states in the baryon number, from which the canonical phase diagram can be determined. 
Even though this dependence is very weak, this method has the potential to discriminate between the chiral and nuclear transition and address the question whether they split, as is expected: in the continuum, chiral symmetry should still be broken in the nuclear phase, resulting in two separate first order transitions at low temperatures. 

\acknowledgments

We thank Aaron von Kamen for his contributions to the Wang-Landau method, and Giuseppe Gagliardi for discussions on the partition function. 
Numerical simulations were performed on the OCuLUS cluster at PC2 (Universität Paderborn).
This work is supported by the Deutsche Forschungsgemeinschaft (DFG) through the Emmy Noether Program under grant No. UN 370/1
and through the CRC-TR 211 'Strong-interaction matter under extreme conditions'– project number 315477589 – TRR 211.

\bibliography{main}

\end{document}